\documentclass[prb,twocolumn,superscriptaddress,showpacs]{revtex4-1}

\pdfoutput=1

\usepackage{times}
\usepackage{amsfonts}
\usepackage{graphicx}
\usepackage{graphics}
\usepackage{amssymb}
\usepackage{layout}
\usepackage{verbatim}
\usepackage{amsfonts,epsfig}
\usepackage{bm}
\usepackage{amsmath}
\usepackage{graphicx}
\usepackage{hyperref}
\usepackage{epsf}
\usepackage{bm}
\usepackage{verbatim}

\newcommand{\up}{\uparrow}
\newcommand{\dn}{\downarrow}
\newcommand \bal {\begin{align} }
\newcommand \eal {\end{align}}

\begin{document}

\title{Interacting topological phases in multiband nanowires}
\author{Roman M.~Lutchyn}
\affiliation{Station Q, Microsoft Research, Santa Barbara, CA 93106-6105}

\author{Matthew P. A. Fisher}
\affiliation{Department of Physics, University of California, Santa Barbara, California 93106}

\date{compiled \today %version A, edited by RL
}
\begin{abstract}
We show that semiconductor nanowires coupled to an s-wave superconductor provide a playground to study effects of interactions between different topological superconducting phases supporting Majorana zero-energy modes. We consider quasi-one dimensional system where the topological phases emerge from different transverse subbands in the nanowire. In a certain parameter space, we show that there is a multicritical point in the phase diagram where the low-energy theory is equivalent to the one describing two coupled Majorana chains. We study effect of interactions as well as symmetry-breaking perturbations on the topological phase diagram in the vicinity of this multicritical point. Our results shed light on the stability of the topological phase around the multicritical point and have important implications for the experiments on Majorana nanowires.

\end{abstract}

\maketitle

\section{Introduction.}

The possibility of realizing Majorana fermions, elusive particles that are
their own anti-particles, in semiconductor nanowires coupled with an s-wave superconductor has attracted a lot of attention recently~\cite{levi'11}. In addition to the intrinsic motivation of finding Majorana particles in nature\cite{Wilczek'09}, the solid-state Majoranas have additional property of fundamental physics interest: Majorana zero-energy modes emerging in topological superconductors obey non-Abelian braiding statistics\cite{read_prb'00, Ivanov_PRL'01,Aliceaetal'10} which can be exploited for quantum computation purposes~\cite{Nayak08}. The prediction of the emergence of Majorana fermions in semiconductor/superconductor heterostructures~\cite{Sau'10, Alicea_PRB10, Lutchyn'10, Oreg'10} has led to much activity aimed at detecting these exotic particles~\cite{Nilsson_PRL08, FuKane'09, LiangFu'10, Law'09, Sau_long} as well as exploiting them for topological quantum computation~\cite{Hassler'10, Sau_TQC, Bonderson'11}.

There is no doubt that Majorana fermions can be realiazed in suitable mean-field models describing realistic physical systems. The existence of Majorana zero-energy modes in these system can be shown theoretically by explicitly solving the corresponding quadratic Hamiltonians~\cite{Sau'10, Fu08, Lutchyn'10, Oreg'10, Cook'11} or invoking topological invariants developed for noninteracting systems~\cite{kitaev'01}. The situation is much more complicated, however, once interactions are included, and there are examples where interactions lead to the breakdown of the classification developed for noninteracting systems~\cite{Fidkowski'10}. In this paper we study effect of interactions on the topological phase diagram using a realistic model which describes multiband nanowires proximity-coupled to an s-wave superconductor. The effective theory naturally emerging in multiband nanowires is equivalent to the model of two coupled Majorana chains. Rather than being spatially dependent, the coupling between Majorana ``chains" is controlled by external parameters such as magnetic field and chemical potential, and can be tuned in a given sample. Within this model, we characterize the effect of interparticle interactions as well as various other perturbations on the topological phase diagram. We find that interactions do not change the phase diagram at the qualitative level but lead to a non-trivial renormalization of the phase boundary.

The paper is organized as follows. In Sec.~\ref{sec:model} we first introduce theoretical model for multi-band Majorana nanowires and discuss topological phase diagram assuming no mixing between different subbands. We identify multicritical points in the phase diagram and derive an effective model describing the multi-critical points. In Sec.~\ref{sec:symbreaking} we consider all band-mixing terms allowed by the symmetry of the model and analyze their effect on the phase diagram, in particular around multicritical points. In Sec.~\ref{sec:interaction} we add interparticle interaction and calculate the topological phase diagram of the system using bosonization and real-space RG techniques. Finally, we conclude in Sec.~\ref{sec:conclusions} with the summary of main results.

\section{Theoretical model.}\label{sec:model}
\begin{figure}
\includegraphics[width=3.4in,angle=0]{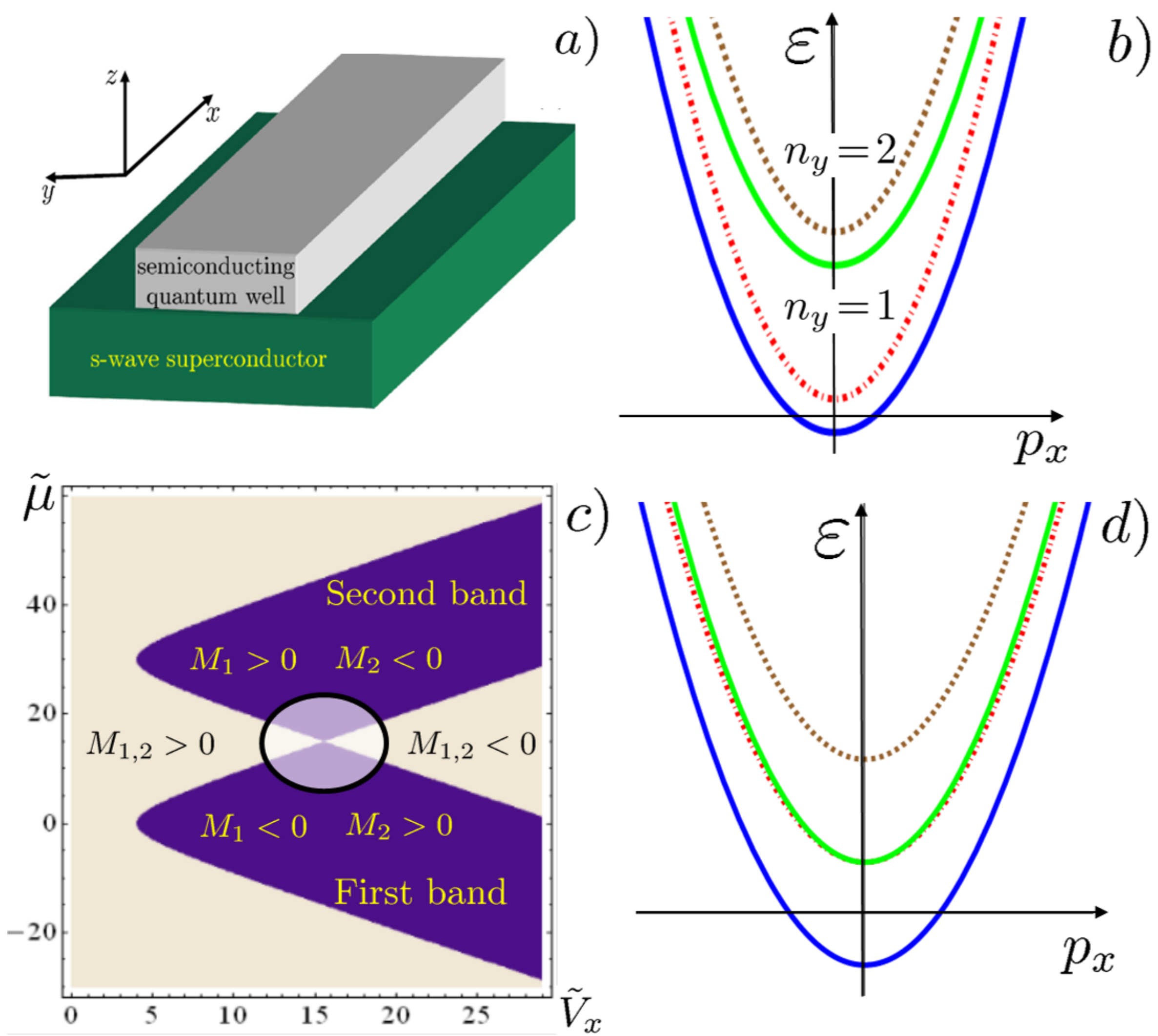}
\caption{(Color online) a) Semiconductor quantum well with dimensions $L_z \ll L_y \ll L_x$  in contact with an s-wave superconductor. b) Single-particle energy spectrum showing lowest two transverse subbands. The combination of the Rashba spin-orbit coupling and magnetic field results in splitting of the spin degeneracy in each subband. c) Phase diagram for two-band topological superconductor assuming fermion parity in each subband is conserved. At the multicritical point the system has enhanced $Z_2 \otimes Z_2$ symmetry and is equivalent to the model of two Majorana chains. Changing of the mass terms $M_a$ corresponds to a topological phase transition and allows one to map out the topological phase diagram. Here tilde denotes rescaled energy $\tilde E\equiv E/m^*\alpha^2$. d) Single-particle energy spectrum at the multicritical point where two bands belonging to different subbands touch, and the band topology changes in a non-trivial way.}
\label{fig:phase_diag}
\end{figure}

The system we consider here consists of a semiconductor quantum well with dimensions $L_z \ll L_y \ll L_x$  in contact with an s-wave superconductor, see Fig.\ref{fig:phase_diag}a. We assume that the confinement along the $z$-axis is very strong so that only the lowest subband with respect to the $z$-axis eigenstates is occupied, whereas the confinement along the $y$-axis is much weaker and only a few $y$-subbands can be populated. Assuming that electrostatic gradient due to the applied gate potential is along $z$-direction, the single-particle Hamiltonian takes the usual form for the 2D semiconductor with the spin-orbit Rashba interaction ($\hbar=1$):
\begin{align}
{\cal H}_{\rm SM}&=\int dx dy \, \psi^\dag_{\lambda}(x,y)\hat H_{\lambda \lambda'}\psi_{\lambda'}(x,y)\label{eq:H0a}\\
H&=-\frac{\partial_x^2+\partial_y^2}{2m^*}-\mu-i\alpha (\sigma_x \partial_y-\sigma_y\partial_x)+V_x\sigma_x, \label{eq:H0b}
\end{align}
where $m^*$, $\alpha$ and $\mu$ are the effective mass, the strength
of spin-orbit interaction and chemical potential,
respectively. The latter can be controlled using the gate electrodes~\cite{Doh'05}.
The last term in Eq.~\eqref{eq:H0b} describes the Zeeman term
due to an applied external magnetic field aligned along the $\hat
x$-axis, $V_x=g_{\rm SM} \mu_B B_x/2$. One can  notice that Hamiltonian~\eqref{eq:H0b} is invariant under transformation $\psi_{\lambda}\rightarrow i \psi_{-\lambda}(-x)$ with $\lambda$ being an electron spin. This transformation corresponds to a $\pi$-rotation around $x$-axis combined with the inversion operation $P\cdot f(x)=f(-x)$:
\begin{align}\label{eq:symmetry}
UH U^{-1}=H \mbox{ \,  where  \, } U=i\sigma_x P.
\end{align}

The proximity-induced superconductivity due to the exchange of electrons between the semiconductor and superconductor leads to the emergence of anomalous correlations in the semiconductor which are captured by the following Hamiltonian:
\begin{align}\label{eq:pairing}
H_{\rm SC}=\int dx dy \left[\Delta_0 \psi_{\up}(x,y)\psi_{\dn}(x,y)+h.c.\right]
\end{align}
We choose the gauge where the induced pair potential $\Delta_0$ is real. If $\Delta_0$ is $x$-independent which is assumed throughout the paper, the Hamiltonian~\eqref{eq:pairing} also respects symmetry~\eqref{eq:symmetry}. Thus, all terms that are derived from the total Hamiltonian $H_T=H+H_{\rm SC}$ should respect the symmetry ~\eqref{eq:symmetry}.

We now construct effective two-band model for the semiconductor nanowire which captures the physics we are interested in. First of all, the Hamiltonian~\eqref{eq:H0b} is separable in $x-y$ coordinates and one can write the field operator as
\begin{align}\label{eq:psi}
\!\!\psi_{\lambda}(x, y)\!=\!\sum_{p_x, n_y=1,2,...}\!\sqrt{\frac{2}{L_y L_x}}\!\sin\left(\frac{\pi n_y y}{L_y}\right)\!e^{ip_x x}\! a_{\lambda, p_x, n_y},
\end{align}
where $a_{\lambda,p_x, n_y}$ is electron annihilation operator in a state
$n_y$ having momentum $p_x$ and spin $\lambda$. Assuming that the confinement energy along $y$-direction $E_{\rm sb}$ is larger than all the relevant energy scales of the Hamiltonian~\eqref{eq:H0b}, one can project the Hamiltonian $H_T$ to the lowest two subbands. Within this approximation, the Hamiltonian of the system reads ($\hbar=1$):
\begin{align}\label{eq:two-banda}
\!\! H_{\rm SM}&\!=\!\sum_{\lambda \lambda'}\!\int_{-L}^{L} dx \! \left[\!c_{\lambda}^\dag\!\left(\!-\frac{\partial^2_x}{2m^*}-\mu\!+\!V_x\sigma_x\!+\!i\alpha\sigma_y \partial_x\!\right)_{\lambda \lambda'}\!c_{\lambda'}\right.\nonumber\\
&\left.\!+\!d_{\lambda}^\dag\!\left(\!-\!\frac{\partial^2_x}{2m^*}\!-\!\mu\!+\!E_{\rm sb}\!+\!V_x\sigma_x\!+\!i\alpha\sigma_y \partial_x\!\right)_{\lambda \lambda'}d_{\lambda'}\!\right]\nonumber\\
H^{(12)}_{\rm so}&=E_{\rm bm} \sum_{\lambda \lambda'}\!\int_{-L}^{L} dx \! \left[c^\dag_{\lambda}\left(i \sigma_x\right)_{\lambda\lambda'}d_{\lambda'}-d^\dag_{\lambda}\left(i\sigma_x\right)_{\lambda\lambda'}c_{\lambda'}\right].\\
H_{\rm P}&=\int_{-L}^L dx \left[ \Delta_0 c_{\uparrow} c_{\downarrow}+ \Delta_0 d_{\uparrow} d_{\downarrow}%+\Delta_{12}(d_{\uparrow}^\dag c_{\downarrow}^\dag+c_{\uparrow}^\dag d_{\downarrow}^\dag)
+h.c.\right].\label{eq:two-bandb}
\end{align}
Here $c_{\lambda}$ and $d_{\lambda}$ represent fermion annihilation operators of the first and second subbands having spin $\lambda$. The spin-orbit band mixing Hamiltonian $H^{(12)}_{\rm so}$
originates from taking the expectation value $\hat p_y$ operator between
different band eigenstates. The energy $E_{\rm
bm}=\int_0^{L_y}dy\frac{2\alpha}{L_y}\sin(\frac{2\pi y}{L_y})\partial_y
\sin(\frac{\pi y}{L})=\frac{8\alpha}{3L_y}$. One can notice that a constant magnetic field $B_x$ does not lead to inter-subband mixing because the corresponding matrix elements vanish identically due to the orthogonality of the wavefunctions corresponding to different subbands. Thus, the two-band approximation is valid even for a large Zeeman splitting $V_x \sim E_{\rm sb}$.

It is implicitly assumed in Eqs.~\eqref{eq:two-banda}-\eqref{eq:two-bandb} that the length of the wire $2L_x\equiv 2L$ is much longer than the effective superconducting coherence length $\xi$ in the semiconductor. For simplicity, we take the induced superconducting pair potential to be the same for the first and second transverse bands.

As shown in Ref.~\onlinecite{Lutchyn_multi}, the system described by the Hamiltonian $H_0=H_{\rm SM}+H_{\rm P}$ realizes a non-trivial topological SC state in a suitable parameter regime. In the weak coupling limit $\Delta_0 \rightarrow 0$, the topological phase emerges when there is an odd number of Fermi surfaces. In this case, the Hamiltonian of the system is adiabatically connected with the one of a spinless p-wave superconductor which is known to host Majorana zero-energy modes at the ends of the nanowire. As shown in Fig.~\ref{fig:phase_diag}b, such a situation is realized when $|V_x|>|\mu|$ and $|V_x|>|\mu-E_{\rm sb}|$ which corresponds to the topological phases originating from the first ($n_y=1$) and second ($n_y=2$) subbands, respectively. Looking at the single-particle band structure as a function of $V_x$, one can notice that there is a special point $V_x\approx E_{\rm sb}/2$ where two bands belonging to $n_y\!=\!1$ and $n_y=2$ subbands touch, see Fig.~\ref{fig:phase_diag}c and d. If the chemical potential is tuned to $\mu=E_{\rm sb}/2$, the single particle band topology changes in a non-trivial way at this point (i.e. the first Chern number defined for two-band Bogoliubov-de Gennes Hamiltonian changes by 2) yielding an interesting phase diagram which is similar to two coupled Majorana chains. In this case, however, instead of originating from physically different chains, Majorana fermions emerge here from different transverse subbands. We first analyze the phase diagram assuming the fermion parity in each subband is preserved, and then discuss how various perturbations including interactions affect the topological phase diagram.

We now provide a qualitative discussion of the topological phase diagram. For the moment, we neglect the band-mixing term $H^{(12)}_{\rm so}$. As shown below, $H^{(12)}_{\rm so}$ drops out from the effective low-energy theory at the multicritical point. Then, the two subbands decouple and one can define fermion parity in each subband, and  one can introduce a $Z_2$ topological invariant $\cal M$ (Majorana number)~\cite{kitaev'01, Lutchyn_multi} for each subband ${\cal M}_1$ and ${\cal M}_2$ where ${\cal M}$ is defined as
\begin{align}\label{eq:Majorana}
{\cal M}={\rm sgn}\left[{\rm Pf} B(0)]{\rm sgn}[{\rm Pf} B\left(\frac {\pi}{a}\right)\right]=\pm 1.
\end{align}
The antisymmetric matrix $B$ in Eq.\eqref{eq:Majorana} represents the Hamiltonian of the system in the Majorana basis~\cite{kitaev'01}. For the special momenta $P=0, \pi/a$ the antisymmetric matrix $B$ can be constructed by the virtue of particle-hole symmetry: $B(P)\!=\!H_{\rm BdG}(P)U$~[\onlinecite{Lutchyn_multi, Ghosh'10}] where unitary matrix $U$ defines particle-hole symmetry of BdG Hamiltonian
\begin{align}
\Theta H_{\rm BdG}(p)\Theta^{-1}=-H_{\rm BdG}(-p)
\end{align}
 with $\Theta=UK$ being the anti-unitary operator. Here $K$ denotes complex conjugation. The change of the topological invariant signals the phase transition with the phase boundary given by $|V_x|\!=\!\sqrt{\mu^2\!+\!\Delta_0^2}$ and $|V_x|=\sqrt{(\mu\!-\!E_{\rm sb})^2\!+\!\Delta_0^2}$ for $n_y=1$ and $n_y=2$ subbands, respectively. At a special point in the phase diagram $\mu=E_{\rm sb}/2$ and $V_x=\sqrt{E_{\rm sb}^2/4+\Delta_0^2}$, two topological phases can coexist and the symmetry group is $Z_2 \otimes Z_2$, see Fig.~\ref{fig:phase_diag}c. Around this point there are four distinct phases: non-topological (no Majorana modes), topological with Majorana fermions originating either from $n_y=1$ or $n_y=2$ subbands, and the last one with two Majorana modes localized on each end, see Fig.~\ref{fig:phase_diag}c. Thus, multiband semiconductor nanowires are interesting both from a fundamental and a practical point of view as they offer a possibility to investigate interaction between various topological phases in a realistic experimental system.

From now on we focus on the multicritical point, and present a simple explanation of the topological phase transition by deriving an effective long wavelength model around it. The topological phase transition requires vanishing of the excitation gap in the system for the reconstruction of the energy spectrum to occur~\cite{read_prb'00}. The quasiparticle excitation gap at the phase boundary vanishes as $E(p)\sim |p|$~[\onlinecite{Aliceaetal'10}], and, thus, one can understand the phase diagram by deriving an effective model in the spirit of $k \cdot p$ perturbation theory. The phase transition between various topological phases can be captured by studying the Dirac-like Hamiltonian with the mass term(s) $M$ changing sign across the phase boundary, see Fig.~\ref{fig:phase_diag}c. We first calculate exact eigenstates around the multicritical point, and then evaluate small corrections due to the deviation of the physical parameters away from this point. Assuming that these terms are small compared to $\mu, \Delta_0, V_x \sim E_{\rm sb}$, we perform the following canonical transformation and project the system to a low energy subspace $\varepsilon \ll E_{\rm sb}$:
\begin{align}\label{eq:canon}
\!c_{\uparrow/\downarrow}\!&\!\approx\!\pm u_- (e^{i\frac \pi 4}\gamma^{(1)}_{R}\!+\!e^{-i\frac \pi 4} \gamma^{(1)}_{L})\!+\!u_+(e^{-i\frac \pi 4}\gamma^{(1)}_{R}\!+\!e^{i\frac \pi 4} \gamma^{(1)}_{L}),%\mbox{ and } c_{\downarrow}=(-u_- \gamma_4+u_+\gamma_4^\dag)
\nonumber\\\\
\!d_{\uparrow/\downarrow}\!&\!\approx\!\mp u_+(e^{i\frac \pi 4} \gamma^{(2)}_{R}\!+\!e^{-i\frac \pi 4}\gamma^{(2)}_{L})\!-\!u_- (e^{-i\frac \pi 4}\gamma^{(2)}_{R}\!+\!e^{i\frac \pi 4} \gamma^{(2)}_{L})\nonumber. % \mbox{ and } d_{\downarrow}=(u_- \gamma_3^\dag-u_+\gamma_3)
%\!d_{\uparrow/\downarrow}\!&\!=\!\mp u_+(\gamma^{(1)}\leftrightarrow \gamma^{(2)})\!-\!u_- (\gamma^{(1)}\leftrightarrow \gamma^{(2)})\nonumber\\ % \mbox{ and } d_{\downarrow}=(u_- \gamma_3^\dag-u_+\gamma_3)
\end{align}
Here $\gamma^{(a)}_{R/L}$ are right/left-moving Majorana operators originating from the first ($a\!=\!1$) and second ($a\!=\!2$) subbands, respectively. In the transformation above we kept only low energy degrees of freedom and neglected high-energy modes. (Note, however, that these high-energy modes are necessary to satisfy canonical anticommutation relations.) The amplitudes $u_{\pm}$ are given by
\begin{equation}
u_{\pm}=\frac{1}{2\sqrt{2}}\frac{\sqrt{E_{\rm sb}^2+4\Delta_0^2\pm E_{\rm sb}\sqrt{E_{\rm sb}^2+4\Delta_0^2}}}{\sqrt{(E_{\rm sb}^2+4\Delta_0^2)}}.\nonumber
\end{equation}
After some algebra, one arrives at the following effective Hamiltonian valid in the vicinity of the multicritical point:
\begin{align}\label{eq:Heff}
\!\!H_0\!\!\approx\!\! \sum_{a=1,2}\!\int_{-L}^{L}\!\! dx\! \left[i\tilde \alpha(\gamma^{(a)}_{L}\partial_x \gamma^{(a)}_{L}\!-\!\gamma^{(a)}_{R}\partial_x \gamma^{(a)}_{R})\!+\!iM_a \gamma^{(a)}_{L}\!\gamma^{(a)}_{R}\!\right],
\end{align}
where $\tilde\alpha\!=\! \alpha \Delta_0/\sqrt{E_{\rm sb}^2+4\Delta_0^2}$ and the mass terms $M_a$ can be written in terms of the deviations from the multicritical point $\delta V_x=V_x -\sqrt{E_{\rm sb }^2/4+\Delta_0^2}$ and $\delta \tilde \mu =E_{\rm sb}(\mu-E_{\rm sb}/2)/\sqrt{E_{\rm sb}^2+4\Delta_0^2}$:
 \begin{align}
 M_{1/2}=-\delta V_x \pm \delta \tilde \mu.
 \end{align}
 %with $\delta \tilde \mu$ being the rescaled chemical potential $\delta \tilde \mu=\delta \mu \frac{E_{\rm sb}}{\sqrt{E_{\rm sb}^2+4\Delta^2}}$.

 The topological phase transitions can be classified in terms of the sign change of the mass terms $M_a$, see Fig.~\ref{fig:phase_diag}c. It is clear that the phase with $M_{1,2}\!>\!0$ is trivial since it is adiabatically connected to $V_x=0$ limit. A sign change of one of the two mass terms corresponds to a topological transition to a phase with a single Majorana mode. Finally, when both mass terms change sign we have two Majorana modes per end. However, the latter state is unstable against perturbations that break fermion parity in the individual chain and ultimately lift the degeneracy by hybridizing the Majorana modes.

\section{Effect of symmetry-breaking terms.}\label{sec:symbreaking}

 Let us consider now effect of the symmetry-breaking terms, which couple Majorana modes $\gamma^{(1)}$ and $\gamma^{(2)}$, on the topological phase diagram. The presence of such terms reduces the $Z_2 \otimes Z_2$ symmetry of the model introduced in Eq.~\eqref{eq:Heff} to $Z_2$ where the two phases corresponds to a different total fermion parity of the system~\cite{kitaev'01}. We now consider all possible $y$-dependent (independent of $x$) fermion bilinear perturbations respecting the symmetry~\eqref{eq:symmetry}:
\begin{align}\label{eq:Hbandmixing}
H_n^{(12)}&=\int_{-L}^{L} dx \left[t_1 (c_{\up}^\dag d_{\up}+c_{\dn}^\dag d_{\dn})+t_2(c_{\up}^\dag d_{\dn}+c_{\dn}^\dag d_{\up})+h.c.\right]\\
H^{(12)}_{\rm a}&\!=\!\!\int_{-L}^{L}dx[\Delta_{12}(d_{\uparrow} c_{\downarrow}\!-\!d_{\downarrow}c_{\uparrow})\!+\!\Delta'_{12}( c_{\uparrow}d_{\uparrow}\!-\!c_{\downarrow}d_{\downarrow})\!+\!h.c.]
\end{align}
In the gauge when $\Delta_0$ is real, $\Delta_{12}$ should also be real which is assumed below whereas the interband tunneling matrix elements $t_1$ and $t_2$ can be in general complex, i.e. $t_i=|t_i|e^{\alpha_i}$. Using Eqs.~\eqref{eq:canon}, one finds that
\begin{align}\label{eq:transformation}
\!\!(c_{\up}^\dag d_{\up}\!+\!c_{\dn}^\dag d_{\dn})\!&\!=\!-\frac{\Delta_0}{\sqrt{4\Delta_0^2\!+\!E_{\rm sb}^2}}\left(\gamma_L^{(1)}\gamma_L^{(2)}\!+\!\gamma_R^{(1)}\gamma_R^{(2)}\right)\\
(c_{\up}^\dag d_{\dn}\!+\!c_{\dn}^\dag d_{\up})&\!=\!-\frac{i\Delta_0}{\sqrt{4\Delta_0^2\!+\!E_{\rm sb}^2}}\left(\gamma_R^{(1)}\gamma_L^{(2)}\!-\!\gamma_L^{(1)}\gamma_R^{(2)}\right)\\
(d_{\up}c_{\dn}-d_{\dn}c_{\up})&=\frac{E_{\rm sb}}{2\sqrt{4\Delta_0^2+E_{\rm sb}^2}}\left(\gamma_L^{(1)}\gamma_L^{(2)}+\gamma_R^{(1)}\gamma_R^{(2)}\right)\nonumber\\
&-\frac i 2 \left(\gamma_R^{(1)}\gamma_L^{(2)}\!-\!\gamma_L^{(1)}\gamma_R^{(2)}\right)\\
(c_{\up}d_{\up}-c_{\dn}d_{\dn})&=\frac{i E_{\rm sb}}{2\sqrt{4\Delta_0^2+E_{\rm sb}^2}}\left(\gamma_R^{(1)}\gamma_L^{(2)}-\gamma_L^{(1)}\gamma_R^{(2)}\right)\nonumber\\
&-\frac 1 2 \left(\gamma_R^{(1)}\gamma_R^{(2)}\!+\!\gamma_L^{(1)}\gamma_L^{(2)}\right)
\end{align}
The bottom term in the above equation is allowed by symmetry~\eqref{eq:symmetry} but invokes triplet pairing which is not generated in the setup considered here. Substituting these results into Eq.\eqref{eq:Hbandmixing}, we arrive at
\begin{align}\label{eq:Hbandmixing1}
H_n^{(12)}=&\int_{-L}^{L} dx \left[ \frac{2i|t_1|\sin \alpha_1 \Delta_0}{\sqrt{4\Delta_0^2\!+\!E_{\rm sb}^2}}\left(\gamma_L^{(1)}\gamma_L^{(2)}\!+\!\gamma_R^{(1)}\gamma_R^{(2)}\right)\right.\nonumber\\
&\left.+\frac{2i|t_2|\cos \alpha_2 \Delta_0}{\sqrt{4\Delta_0^2\!+\!E_{\rm sb}^2}}\left(\gamma_L^{(1)}\gamma_R^{(2)}\!-\!\gamma_R^{(1)}\gamma_L^{(2)}\right)\right],\\
H^{(12)}_{\rm a}=& - i\Delta_{12}\int_{-L}^{L} dx [\gamma^{(1)}_{R}\gamma^{(2)}_{L}-\gamma^{(1)}_{L}\gamma^{(2)}_{R}].
\end{align}
Thus, at the level of the effective low-energy theory, there are only two types of allowed symmetry-breaking terms which couple Majorana fermions in different subbands, and modify the phase diagram at the qualitative level. The physical origin of these terms requires breaking translational symmetry along $y$-axis. (Note that the spin-orbit band-mixing Hamiltonian $H_{\rm so}^{(12)}$ does not contribute to the effective long-wavelength model since $H_{\rm so}^{(12)}$ corresponds to $t_2$-perturbation with $\alpha_2=\pi/2$.) In an experimental system, the semiconductor/superconductor interface can break translational invariance along $y$-axis. Alternatively, one may apply $y$-dependent electrostatic potential to generate band-mixing terms.

We can now analyze the full Hamiltonian $H_0+H_{12}$, where $H_{12}$ reads
\begin{align}
H_{12}&=i\lambda_1\int_{-L}^{L} dx \left[\gamma^{(1)}_{R}\gamma^{(2)}_{L}-\gamma^{(1)}_{L}\gamma^{(2)}_{R}\right]\\
&+i\lambda_2 \int_{-L}^{L} dx \left[\gamma_L^{(1)}\gamma_L^{(2)}\!+\!\gamma_R^{(1)}\gamma_R^{(2)}\right],
\end{align}
and understand effect of the symmetry-breaking perturbations. The modified phase boundary can be obtained by looking at closing of excitation gap. The lowest branch of the excitation spectrum is given by
\begin{widetext}
\begin{align}
E(p)=\sqrt{\delta V_x^2+\delta \mu^2+\tilde\alpha^2 p_x^2+\lambda_1^2+\lambda_2^2-2\sqrt{\tilde \alpha^2 p_x^2 \lambda_2^2+\delta V_x^2(\delta \mu^2+\lambda_1^2+\lambda_2^2)}},
\end{align}
\end{widetext}
and vanishing of the excitation gap at $p_x=0$ defines a new phase boundary:
\begin{align}\label{eq:newphase_bound}
|\delta V_x|=\sqrt{\delta \mu^2+\lambda_1^2+\lambda_2^2}.
\end{align}
Thus, at $\delta \mu=0$ there is a window $-\sqrt{\lambda_1^2+\lambda_2^2}<\delta V_x < \sqrt{\lambda_1^2+\lambda_2^2}$ where topological phases from different subbands coexist corroborating the results of Ref.~\onlinecite{Lutchyn_multi}. We note that this scenario occurs as long as the Hamiltonian~\eqref{eq:H0b} respects the symmetry~\eqref{eq:symmetry}. The alternative scenario where two topologically trivial phases hybridize is forbidden by the symmetry \eqref{eq:symmetry}.
As emphasized in Ref.~[\onlinecite{Lutchyn_multi}], this effect is particularly important for experimental realization of the topological superconducting phase in semiconductor nanowires where chemical potential fluctuations pose serious constraint. Indeed, the topological phase around this region is to a large extent robust against chemical potential fluctuations which now have to be $\delta \mu \sim E_{\rm sb}$ to cause the transition into the nontopological state.

One can notice that at $\delta \mu=0$ and $\lambda_1=0$ the quasiparticle excitation gap also closes at a finite momentum $|p_x|=\sqrt{\lambda_2^2-\delta V_x^2}/\tilde\alpha$. However, closing of the gap in this case does not correspond to a topological transition which should be accompanied by a change of the topological invariant. As follows from Eq.~\eqref{eq:Majorana}, the topological invariant can change only at the particle-hole symmetric points $p_x=0, \pi/a$. Therefore, only gap closing at these point corresponds to a topological phase transition.

%Since the  which in 1D can change only at

\begin{figure}
\includegraphics[width=3.4in,angle=0]{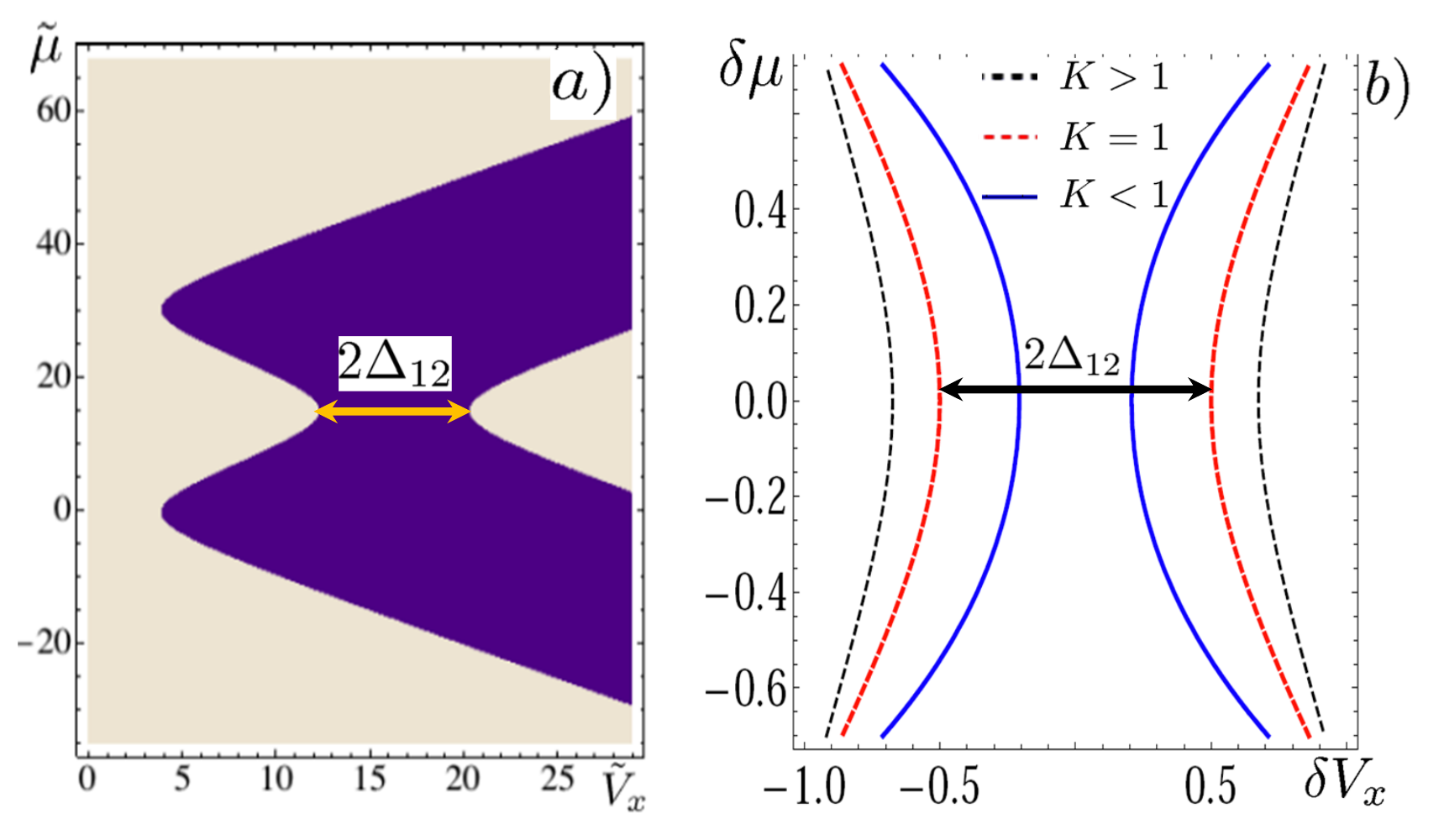}
\caption{(Color online) a) Topological phase diagram for two band semiconductor model in the presence interband superconducting pairing (i.e. $y_1=\Delta_{12}$ and $y_2=0$) which breaks fermion parity in the individual subband and leads to the hybridization between the Majorana modes originating from $n_y\!=\!1$ and $n_y\!=\!2$ subbands. As a result, the enhanced $Z_2\otimes Z_2$ symmetry at the multicritical point is broken down to $Z_2$, compare with Fig~\ref{fig:phase_diag}c. b) Shift of the topological phase boundary caused by interactions. Repulsive/attractive interaction lead to the suppression/enhansement of the topological phase at the multicritical point.}
\label{fig:interaction}
\end{figure}

\section{Interaction effects.}\label{sec:interaction}

We now study the effect of interactions on the topological superconducting phase. For simplicity, we consider short-range interactions given by the Hamiltonian
\begin{align}\label{eq:interaction}
\!\!H_{\rm int}\!&=\!U\!\! \sum_{\lambda,\lambda'\!=\!\uparrow,\downarrow}\!\!\! \int_{-L}^{L}\! dx\! :\!(c_{\lambda}^\dag c_{\lambda}\!+\!d_{\lambda}^\dag d_{\lambda})\!::\!(c_{\lambda'}^\dag c_{\lambda'}\!+\!d_{\lambda'}^\dag d_{\lambda'})\!:\\
&\!\approx\!-\tilde U \!\int_{-L}^{L}\! dx\! :\!(\gamma^{(1)}_L \gamma^{(1)}_R\!-\!\gamma^{(2)}_L \! \gamma^{(2)}_R)\!::\!(\gamma^{(1)}_L \!\gamma^{(1)}_R\!-\!\gamma^{(2)}_L \!\gamma^{(2)}_R)\!,\nonumber
\end{align}
where $\tilde U=U \frac{E_{\rm sb}^2}{E_{\rm sb}^2+4\Delta_0^2}$. In the last line of Eq.~\eqref{eq:interaction} we have projected the Hamiltonian to the low energy subspace. Combining all the terms, the full Hamiltonian for the interacting system becomes $H=H_0+H_{12}+H_{\rm int}$. One can see that at the multicritical point (assuming $\lambda_1=0$ and $\lambda_2=0$), the Hamiltonian $H$ maps onto Thirring model which can be solved exactly using bosonization. Note that the interaction term \eqref{eq:interaction} does not break fermion parity in the individual chain, and preserves $Z_2 \otimes Z_2$ symmetry present at the multicritical point. The effect of deviations from the multicritical point as well as effect of symmetry-breaking terms can be included in the model perturbatively assuming that these terms are small to begin with. We proceed using standard bosonization~\cite{Giamarchi_book} by first introducing left and right-moving Dirac fermions $\psi_{R/L}=(\gamma^{(1)}_{R/L}-i\gamma^{(2)}_{R/L})/\sqrt 2$, and then rewriting them in terms of the bosonic fields $\varphi$ and $\theta$: $\psi_{R/L} \sim e^{i(\varphi \pm \theta)}/\sqrt{2\pi a}$ %as $\psi_{R/L}=\frac{U_{R/L}}{\sqrt{2\pi a}}e^{-i[\pm \Phi-\Theta]}$
. The bosonized Hamiltonian $H$ becomes
\begin{align}\label{eq:TM0}
\!H\!\!=\!\!&\!\!\int_{-L}^{L}\!\! d x\!\! \left[\frac{v}{2\pi}\left(\!K(\partial_x \varphi)^2\!+\!K^{-1}\! (\partial_x \theta)^2\!\right)-\!\frac{2\delta V_x}{2\pi a}\!\sin 2\theta\right.\nonumber\\
\!&\left.\!+\!\frac{2\delta \mu}{2\pi a}\!\sin 2\varphi\!-\!\frac{2\lambda_1}{2 \pi a}\!\cos2\varphi\!-\frac{2\lambda_2}{v}\partial_x\theta\right]\!,
\end{align}
where $v$, $K$ are related to the microscopic parameters of the model:
%\begin{align}
$v\!=\!\tilde \alpha \sqrt{1\!-\!\left(\tilde U/\pi\tilde \alpha\right)^2}$,  $K=\sqrt{(1\!-\!\frac{\tilde U}{\pi\tilde \alpha})/(1\!+\!\frac{\tilde U}{\pi\tilde \alpha})}$,
%\end{align}
Here $a$ is a cutoff in the problem, which is related to the momentum bandwidth $\Lambda \!\sim\! 1/a$~[\onlinecite{Giamarchi_book}]. To simplify Hamiltonian~\eqref{eq:TM0} we shift $2\varphi\rightarrow 2\varphi-\tan^{-1}(\lambda_1/\delta \mu)$ and introduce dimensionless parameters
$y_1\!=\!\delta V_x a/v$, and $y_2\!=\!\sqrt{\lambda_1^2\!+\!\delta {\tilde \mu}^2}a/v$. Then, the Hamiltonian becomes~\eqref{eq:TM0}
\begin{align}\label{eq:TM}
\!H_{\rm eff}\!\!&=\frac{v}{2\pi}\!\int_{-L}^{L}\!\! d x\!\! \left[\!K(\partial_x \varphi)^2\!+\!K^{-1}\! (\partial_x \theta)^2\!-\!\frac{2y_1}{a^2}\!\sin 2\theta\!\right.\nonumber\\
&\left.+\!\frac{2y_2}{a^2}\!\sin 2\varphi\!-\frac{2\lambda_2}{v}\partial_x \theta \right]\!.
\end{align}
The effective Hamiltonian $H_{\rm eff}$ describes the physics around the multicritical point in the presence of interactions and various symmetry-breaking perturbations. In the rest of the paper, we analyze Eq.~\eqref{eq:TM} and compute corrections to the topological phase boundary due to the presence of the interactions.

\subsection{Phase diagram for $\lambda_2=0$.}

It is instructive to first analyze the phase diagram with $\lambda_2=0$. In this case, the Hamiltonian~\eqref{eq:TM} appears in other systems such as classical two-dimensional XY model in a magnetic field and weakly coupled Heisenberg chains, see, e.g. Refs.~\onlinecite{Lecheminant'02, Jose'77, Fertig'02}, and one can develop some intuition based on these analogies.

Let us investigate how interactions modify the topological phase diagram of the system and study effects of the competing relevant operators on the non-trivial criticality. %As mentioned above at $K=1$ there is a gapless mode when $|y_1|=|y_2|$.
Assuming that $y_{1,2}\rightarrow 0$ and small interactions $|K-1|\ll 1$, one can obtain flow equations using a perturbative real-space RG approach~\cite{Jose'77}:
\begin{align}\label{eq:RG}
\frac{d y_1(l)}{dl}&=(2-K)y_1(l),\\
\frac{d y_2(l)}{dl}&=(2-K^{-1})y_2(l),\\
\frac{d\ln K}{dl}&=K^{-1}y_2^2-K y_1^2,
\end{align}
where $l=\ln[a/a_0]$ is the flow parameter with $a_0$ being the initial value of the cutoff. Above RG equations reflect duality in the model: $\varphi\leftrightarrow \theta$ at $K \leftrightarrow K^{-1}$ and $y_1 \leftrightarrow y_2$. One can see that in the vicinity of $K=1$, where we have a non-trivial critical point, both mass terms are relevant and flow to strong coupling under RG, i.e. each perturbation acting separately would yield a massive field theory. However, given that $y_1$ and $y_2$ couple to dual field operators corresponding to charge-density-wave pairing and Cooper-pairing, respectively, and drive the system to different ground states, the interplay between them gives rise to a second-order phase transition at the intermediate coupling. Indeed, this is what happens at the self-dual point $K\!=\!1$. Away from this exactly solvable point, some intuition can be obtained by invoking scaling arguments. Using the analogy with 2D classical theory, the scaling theory of our zero-temperature 1D problem can be easily formulated. Since the critical theory should be invariant under the rescaling, the singular part of the energy density satisfies the following scaling relations:
\begin{align}
f_s[y_1,y_2]=\!e^{-2l} f_s[e^{(2-K)l} y_1,e^{(2-K^{-1})l}y_2].
\end{align}
The scale $l$ can be fixed by requiring $e^{(2-K^{-1})l^*}y_2=1$. Substituting $l^*$ into the energy density, one finds
\begin{align}
f_s[y_1,y_2]=\!y_2^{\frac{2}{2-K^{-1}}} f_s\left[y_2^{-\frac{2-K}{2-K^{-1}}} y_1,1\right].
\end{align}
The phase transition in this system occurs when $y_2^{-\frac{2-K}{2-K^{-1}}} y_1\sim 1$, and, thus, the new phase boundary is given by
\begin{align}\label{eq:phase_boundary}
|\delta V_{x}|= \left(\lambda_{1}^2+\delta\mu^2\right)^{\frac 1 2 \frac{2-K}{2-K^{-1}}}.
\end{align}
This is one of the main results of the paper: the presence of weak interactions does not destroy the multicritical point but leads to a non-trivial renormalization of the phase boundary which now depends on the Luttinger liquid parameter $K$. One can easily see that Eq.~\eqref{eq:phase_boundary} is consistent with the noninteracting ($K=1$) result discussed earlier. In the case of a repulsive interaction ($K\!<\!1$), the topological region at the multicritical point shrinks, see Fig.~\ref{fig:interaction}b. This result can be understood as a competition between the repulsive interaction and proximity-induced superconductivity, and in this regard, our conclusions are consistent with those of Refs.~\onlinecite{Loss'11, Sela'11, Stoudenmire'11}. On the contrary, attractive interaction ($K\!>\!1$) stabilizes the topological phase by expanding the area occupied by the topological phase, see Fig.~\ref{fig:interaction}b.

\subsection{Phase diagram for $\lambda_2 \neq 0$.}

 Let us now consider effect of the $\lambda_2$-term  on the topological phase diagram. It is helpful to first make a transformation $2\theta\rightarrow 2\theta +2\delta_Q x$ where $\delta_Q=K\lambda_2/v$. First of all, such a transformation eliminates $\partial_x \theta$ in Eq.\eqref{eq:TM} and leads to spatially oscillating $y_1$ term. When $y_2=0$ the Hamiltonian~\eqref{eq:TM} corresponds to well-known Pokrovsky-Talapov model~\cite{Pokrovsky'79} which exhibits commensurate-incommensurate transition.
 If $\delta_Q a \gg 1$, the $\sin(2\theta +2\delta_Q x)$ is quickly oscillating and averages out to zero which reflects competition of $y_1$ and $\lambda_2$ terms.
 As a result of this competition, the RG flow for $y_1$ has to be cutoff when $2\delta_Q(l) a \sim 1$. To the lowest order in $y_1$, $y_2$ and $\delta_Q$, the RG flow of $\delta_Q$ is given by:
\begin{align}
\frac{d\delta_Q}{dl}=\delta_Q.
\end{align}
Since all perturbations $y_1$, $y_2$ and $\delta_Q$ are relevant, they flow to strong coupling under RG. If $y_1(l)$ reaches strong coupling before $\delta_Q(l)a$ becomes order one,
the phase boundary can be calculated as before and is given by Eq.~\eqref{eq:phase_boundary}. The condition $y_1(l^*)\sim 1 $ defines a new length scale $l^*$. This scenario is self-consistent as long as $2\delta_Q(l^*)a \ll 1$ which translates into the requirement that $\delta_Q(0)a \ll y_1(0)^{\frac{1}{2-K}}$. In the opposite regime, the $y_1$-term does not reach strong coupling whereas the RG flow of $y_2$-term is unaffected by the $\lambda_2$ terms, and, thus, the system flows to an ordered phase determined by $y_2\cos (2\varphi)$. If $y_2$ is exactly zero, the system is massless which corresponds to an incommensurate phase in the context of C-IC transition. For $K=1$ this transition corresponds to closing of the gap at a finite momentum which occurs when $|\lambda_2| > |\delta V_x|$, in agreement with the arguments presented above. As discussed in Sec.~\ref{sec:symbreaking}, closing of the gap at finite momentum is not associated with a change of the topological invariant and, thus, C-IC transition in the bosonic language does not correspond to a topological transition.

\section{Conclusions.}\label{sec:conclusions}

To conclude, we have studied interacting topological superconducting phases using a realistic model corresponding to the semiconductor nanowire in the limit of multiband occupancy in contact with an s-wave superconductor. In the vicinity of the multicritical point in the phase diagram, the model considered here is equivalent to the one describing two coupled Majorana chains, and we have characterized the effect of interactions as well as other perturbations on the topological phase diagram. We find that moderate interactions do not affect the phase diagram qualitatively but lead to nontrivial quantitative changes in the phase boundary. Our results characterize the stability of the topological phase around the multicritical point against interactions and have important implications for the experiments on Majorana nanowires. Our results for the topological phase diagram around multicritical points are generic and should be applicable to situations when there is a larger number of channels occupied in the nanowire ({\it i.e.} more than two) as considered recently in Ref.\onlinecite{tudor'11}.

\section{Acknowledgements}

We would like to thank J. Alicea, F.~Essler, P. Fendley, E. Fradkin, L. Fidkowski, and C.~Nayak for enlightening discussions. We thank the Aspen Center for Physics for hospitality and support under the NSF grant \#1066293. We gratefully acknowledge support from the National
Science Foundation through grant DMR-1101912 (M.P.A.F.).

%\begin{thebibliography}
%\bibliographystyle{apsrev}
%\bibliography{TQC_ref}
%\end{thebibliography}


\end{document}